\documentclass[pra,preprint,aps]{revtex4}
\usepackage{graphics}

\begin{document}

\title{Strange Attractors in Multipath propagation: Detection and characterisation.}

\author{C. Tannous}
\email{tannous@univ-brest.fr}
\altaffiliation{Present address: Laboratoire de Magnétisme de Bretagne, UPRES A CNRS 6135,
Université de Bretagne Occidentale, BP: 809 Brest CEDEX, 29285 FRANCE}
\author{R. Davies}
\affiliation{Alberta Government Telephones Calgary, Alberta, Canada T2G 4Y5}
\author{A. Angus}
\affiliation{NovAtel Communications Calgary, Alberta, Canada T2E 7V8}

\date{March 16, 2001}

\begin{abstract}

Multipath propagation of radio waves in indoor/outdoor environments shows a highly irregular behavior
 as a function of time. Typical modeling of this phenomenon assumes the received signal is a stochastic
 process composed of the superposition of various altered replicas of the transmitted one, their amplitudes
 and phases being drawn from specific probability densities. We set out to explore the hypothesis of the 
presence of deterministic chaos in signals propagating inside various buildings at the University of Calgary.
 The correlation dimension versus embedding dimension saturates to a value between 3 and 4 for various antenna 
polarizations. The full Liapunov spectrum calculated contains two positive exponents and yields through the 
Kaplan-Yorke conjecture the same dimension obtained from the correlation sum. The presence of strange attractors 
in multipath propagation hints to better ways to predict the behaviour of the signal and better methods to counter
 the effects of interference. The use of Neural Networks in non linear prediction will be illustrated in 
an example and potential applications of same will be highlighted.

\pacs{PACS numbers: 05.45.+b,89.70.+c}

\end{abstract}

\maketitle

\section{Introduction}

Multipath propagation of radio waves in indoor or outdoor environments shows a highly irregular behavior as a
 function of time [1]. The characterization of radio channels in mobile or in building propagation is important
 for addressing issues of design, coding, modulation and equalization techniques tailored specifically to combat 
time and frequency dispersion effects.\\
Irregular behavior of the received signal has prompted researchers in the past to model the channel with stochastic
 processes. One of the earliest linear models in this vein is the Turin et al. model [2] in which the impulse response 
of the channel is written as a superposition of replicas of the transmitted signal delayed and having altered amplitudes and phases.\\
A number of models exist differing in the numbers of replicas of the signal or in the type of probability distributions from which the amplitudes and phases are drawn. Also, different stochastic processes are used in the generation of delay times of received replicas. The popular choice for the amplitude probability density functions (PDF) are Rayleigh or Rice PDFs depending on whether a weak or strong line of sight propagation exists; nevertheless other PDFs have been used such the lognormal, 
Nakagami-m or uniform. The phases ought to be drawn from PDFs compatible with the ones selected for the corresponding 
amplitudes; nevertheless the most popular choice found in the literature is the uniform $[0 - 2 \pi]$ distribution. Delay 
times are usually extracted from either stationary of stationary Poisson processes although in some cases the Weibull PDF is used.

\section{A HYPOTHESIS OF DETERMINISTIC CHAOS}

Although an assumption of stochastic behavior in mobile or indoor propagation is ubiquitous, in the present work we 
set out to explore the hypothesis that the indoor communication channel displays deterministically chaotic behavior.
 This question is important in many respects and the tools to answer it readily exist. These tools are based on the 
determination of the correlation dimension of the strange attractor associated with the multipath profile considered 
as a real valued time-series x(t). Using delay coordinates [3] one forms the m-dimensional delay vector ${\bf X}(t) = [x(t),x(t-T) ...x(t-(m-1)T]$ with delay $T$ and computes the correlation sum C(r) which is the ratio of the number of pairs of delay 
vectors (the distance between which is less than r) to the total number of pairs. From this, the correlation dimension 
$\nu$ is defined as the logarithmic slope of C(r) versus r for small r. For a true stochastic process, $\nu$ increases
 with m without showing any saturation. In contrast, for deterministic chaos, $\nu$ saturates at a value, the next integer
 greater than which, represent the minimum number of non-linear recursion or differential equations from which it originates.
 If the profile turns out to be deterministic, the modulation, coding and detection/demodulation techniques ought to be adapted 
accordingly in order to account for this fact; otherwise one has to rely upon techniques capable of handling stochastic 
signals. Let us illustrate this by an analysis of multipath measurements we have made in an indoor environment.

\section{EXPERIMENT AND CORRELATION DIMENSION ANALYSIS}

The propagation environment from which the data are collected are hallways in the Engineering Building at The University 
of Calgary [4]. The transmitted power was 10 dBm fed into a half wave dipole antenna with a matching balun. The receiving
 antenna was a cross-polarized dipole array. The co-polarized antennas (CPA) and crosspolarized antennas (XPA) profiles
 referred to in Figure 1 are from a single measurement run and points from both profiles were obtained in coincident pairs.
The terms CPA and XPA simply refer to the relative state of polarisation between the transmit and receive antennas.
 The receiving hardware was specially developed to measure diversity characteristics and gives an accurate reference 
between distance from transmitting to receiving antennas and received signal strength. The same measurement procedure 
was employed as for the arbitrarily polarized data set. We have estimated the correlation dimension for the sets of 
data: Arbitrarily Polarized Antennas (APA, 6000 data points), co-polarized (CPA, 3800 points) and Cross-polarized 
antennas (XPA, 3800 points) by the box-counting method of Grassberger and Procaccia [3]. There are a number of 
limitations and potential pitfalls with correlation dimension estimation that have been discussed by various authors
 [5]. In addition, the number of operations it takes to estimate C(r) is $O(N^{2})$ where $N$ is the number of collected 
experimental points. Recently Theiler [5] devised a powerful box-assisted correlation sum algorithm based on a 
lexicographical ordering of the boxes covering the attractor, reducing the number of operations to $O(Nlog(N))$ 
and incorporating several test procedures aimed at avoiding the previous pitfalls. Before we used Theiler's
 algorithm, we made some preliminary tests against well known cases. We generated uniform random numbers, 
Gaussian random numbers, and numbers z(n) according to the logistic one-dimensional map at the onset of Chaos: 
z(n+1) = a \hspace{0.5mm} z(n)(1-z(n)) with a = 3.5699456 and in the fully developed Chaotic regime at a=4. In the first two 
cases we found $\nu$ approximately equal to m as expected in purely stochastic series whereas $\nu$ saturated 
respectively at 0.48 and 0.98 (we used 2000 points only) for the logistic map indicating the presence of 
a low dimensional attractor and deterministic Chaos (the exact correlation dimensions for the logistic map
 is 0.5 at a=3.5699456 and 1 for a=4.). We tested as well the H\'{e}non two dimensional map, the Lorenz three 
dimensional system of non-linear differential equations, the R\"{o}ssler three and four dimensional systems as
 well as an infinite dimensional system, the delay-differential Mackey-Glass equation whose attractor
 dimension is tunable with the delay time. The results we found for the various correlation dimensions 
agreed with all the results known in the literature to within a few percents. Then we went ahead and examined 
the  $\nu$ vs. m curves for the three sets of experimental data along with a set of 6000 Rayleigh and band-limited
 Rayleigh distributed numbers which constitute prototypic received envelopes. Our results, in Figure 1 show
 that, while $\nu \sim m$ for the pure Rayleigh case (with a slope equal to one), and $\nu \propto m$ for the band 
limited Rayleigh case (with a slope smaller than one) the $\nu$ vs. m curves for the three examined experimental
 sets of data start linearly with m then show saturation indicating the presence of a low dimensional 
attractor (whose dimension is about 4 for CPA and XPA data whereas it is slightly above 4 for the APA 
situation). This finding is in line with the Ruelle criterion [6] that sets an upper bound on the possible
 correlation dimension one can get from any algorithm of the Grassberger-Procaccia type. This upper bound 
is set by the available number of data points $N$ in the time series. The dimension that can be detected should 
be much smaller than $2 \hspace{2mm} log_{10}(N)$. Since we used 3800 and 6000 points respectively, the upper bound for the
 detectable correlation dimension in our case is about 7.16 to 7.56. We respect this bound since our correlation 
dimensions are around 4. Nevertheless the presence of Chaos is going to be confirmed through another route, the 
spectrum of the Liapunov exponents that will be discusssed next.

\section{SPECTRUM OF THE LIAPUNOV EXPONENTS}

The spectrum of Liapunov exponents is very important in the study of dynamical systems. If the largest exponent is positive,
 this is a very strong indication for the presence of Chaos in the time series originating from the dynamical system. The 
reciprocal of this exponent is the average prediction time of the series and the sum of all the positive exponents (if more
 than one is detected like in hyper-chaotic systems such as the R\"{o}ssler four dimensional system of non-linear differential
 equations or the large delay Mackey-Glass equation) is the Kolmogorov entropy rate of the system. The latter gives a 
quantitative idea about the information processes going on in the dynamical system. In addition, with the Kaplan-Yorke 
conjecture, the full spectrum gives the Hausdorff dimension of the strange attractor governing the long time evolution 
of the dynamical system. We have calculated the Liapunov exponents of the data with four different methods. Firstly,
 we determined the largest exponent $\lambda_{max}$ from the exponential separation of initially close points on the attractor 
and averaging over several thousand iterations. Second, we determined the largest Liapunov exponent from the correlation
 sum with the help of the relation $C(r) \sim r^{\nu} exp(-m T \lambda_{max})$ valid for large values of the embedding dimension
 $m$ and small values of r. Finally, we determined all exponents with two different methods: the Eckmann et al. method [7] 
and the Brown et al.'s [8]. Our results for the spectrum of exponents is shown in figures 2 and 3. We tried several
 embedding delay times T and several approximation degrees for the tangent mapping polynomial (as allowed in the Brown
 et al. [8] algorithm ) without observing major changes in the spectrum. Several time series (Logistic map, H\'{e}non, 
Lorenz, R\"{o}ssler and Mackey-Glass) were tested for the sake of comparison to results obtained with the experimental 
data. In addition, the Liapunov exponents saturate smoothly as they should for large embedding dimension. 
Then we applied the Kaplan-Yorke conjecture to get the dimension of the attractor: Using the following typical 
numbers we obtained for the exponents $\lambda_{max}=\lambda_{1}=18.06, \lambda_{2}=1.88, \lambda_{3}=-8.85, \lambda_{4} =-24.94,
 \lambda_{5}=-68.80$ and using the formula:

\begin{equation}
D=j+\frac{\sum_{i}\lambda_{i}}{|\lambda_{j+1}|}
\end{equation}

where the summation is over i=1,2...j. The $\lambda_{i}$'s are ordered in a way such that they decrease as i increases. 
We determine j from the conditions $\sum_{i} \lambda_{i} > 0 \hspace{2mm} \mbox{and:} \hspace{2mm} \lambda_{j+1} + \sum_{i}\lambda_{i} < 0$. We get j=3 and
 D=3.44 for the strange attractor dimension (called its Liapunov dimension). The total sum of the Liapunov exponents
 is negative ( equal to -82.65) as it should be for dissipative systems with a strange attractor. The value of
 the attractor dimension will be confirmed from the spectrum of singularities or the mutifractal spectrum in the next section.

\section{MULTIFRACTAL SPECTRUM}

The generalized dimension may be used to characterize non-uniform fractals for which there are different scaling 
exponents for different aspects of the fractal, so-called multifractals. For these, there are two scaling exponents, 
one generally called $\tau$, for the support of the fractal, and one called q, for the measure of bulk of the fractal. 
In general, $\tau(q)= (q-1)D_q$, where $D_q$ is the generalized dimension. Multifractals have been employed to characterize
 multiplicative random processes, turbulence, electrical discharge, diffusion-limited aggregation, and viscous fingering [9].
 Multifractals have this in common: there is a non-uniform measure (growth rate, probability, mass) on a fractal support.
 Besides the exponents, $\tau$ and q, and the generalized dimension $D_q$, there is another method for characterizing multifractals.
 This depends upon the use of the mass exponent $\alpha$, and the multifractal spectrum, $f(\alpha)$ [9]. A graph of the multifractal
 spectrum explicitly shows the fractal dimension, f, of the points in the fractal with the mass exponent (or scaling index), $\alpha$.
 We have estimated $D_q$ by use of the generalised moments of the correlation sum with a window chosen carefully enough to avoid
 temporal correlation effects. We have developed a program that computes the generalized correlation sum using a box-assisted method.
 Our program is based on one written by Theiler [5]. Several modifications had to be made to the straightforward box-assisted 
correlation sum method. In addition, our program allows for logarithmic scaling with the distance parameter r. From a log-log
 graph of the generalized correlation sums, appropriate scaling regions can be identified, for each order, q. Least-squares 
fits to these scaling regions yields a sequence of generalized correlation dimensions, $D_q$, for values of q between $\pm \infty$. 
We have found computation of $D_q$ for integers in the interval [-10,10] and $D_{-\infty}$ and $D_{+\infty}$ to be sufficient.
 From the $D_q$, we 
calculate the $\tau(q)= (q-1)D_q$. We then perform a Legendre transform to obtain the $f(\alpha)$ curve. We do this by first 
fitting a smooth curve (a hyperbola was considered to be adequate) to the $\tau(q)$ curve. With an analytic expression for 
the $\tau(q)$ curve, we can compute the Legendre transform in closed form. The domain of $f(\alpha)$ may be found from $D_{-\infty}$ and $D_{+\infty}$; we assume that $\alpha$ is confined to this region, and that $f(\alpha)$ is 0 at these points. The values of 
$f(\alpha)$ for $D_{-\infty} \leq \alpha  \leq D_{+\infty}$ are calculated as min[${q\alpha-\tau(q)}$], the minimum being taken
 over q. We found that this procedure, 
although complex, corrects for the known numerical sensitivities of the Legendre transform. We checked that our method for 
obtaining $f(\alpha)$ gave the same results as those found in the literature for the Logistic map and the strongly 
dissipative circle map [9]. The $f(\alpha)$ curve for the multipath data is shown in Figure 4. It may be seen that 
the peak value of $f(\alpha)$, corresponding to the box-counting dimension at q=0, is about 3.7. This is consistent 
with our above findings from the correlation dimension and the full spectrum of Liapunov exponents. Our further 
research in this area concerns the prediction of the received signal intensity, from our above hypothesis of 
the presence of deterministic chaos.

\section{NON LINEAR PREDICTION}

We applied the above findings to the non linear prediction of multipath profiles considering that each point 
on the envelope of the measured signal is a function of some number of past points in the series, deviating 
from the traditional wave superposition approach. More precisely, we write:

\begin{equation}
y(n+1)=F[y(n),y(n- 1),y(n-2),y(n-3) ...y(n-m+1)]
\end{equation}

with F an $m$ dimensional map and y(n) the value of the signal x(t) sampled at timestep n. Expressing F as a
 sum of sigmoidal and linear functions [10] we determine the unknown weights through the Marquardt least 
squares minimisation method [11] in order to achieve the best fit to the data. Our results comparing 
the onestep prediction to the multipath data analysed above are displayed in figure 5. The goodness of fit
 between the measured and predicted envelopes for a map dimension m=5 is another indication of the soundness
 of the approach. This is confirmed in Figure 6 where we display the normalised prediction error versus the
 embedding dimension. A minimum is observed in the prediction error for an embedding equal to 5 or 6. 
In Figure 6, we started always from the same initial weights and let the system run through 150 iterations 
searching for the least squares minimum for 200 data points and later on making 300 one step ahead predictions.
 For embedding dimension larger or equal to 7 the minimisation procedure stopped because of the presence of 
zero-pivot in the least-squares matrices. The presence of a minimum in the prediction error around an embedding 
equal to 5 or 6 complies again with the value of embedding dimension used previously in the correlation dimension
 analysis, the Liapunov spectrum and the Kaplan-Yorke conjecture.\\

\section{DISCUSSION}

We stress that although the profiles we examined were found to be chaotic in all three experimental configurations
 with confirmations from the Liapunov spectrum and non linear prediction studies, indicating that we would be able
 to describe our data with a set of at most 5 non-linear differential or algebraic equations, investigation in 
other propagation situations is needed. Nevertheless, in our investigations we observed a significant amount of 
consistency between the various methods of detecting Chaos and characterising it using embedding dimensions $m$ 
beyond the minimum  $m_{min}$ required by the Takens theorem ($m_{min} > 2d+1$, where d is the dimension of the strange 
attractor). Assuming the hypothesis of the presence of Chaos in a given multipath profile is firmly established, 
many avenues become possible. For instance, one might consider devising ways for controlling the signal propagation 
by altering slightly some accessible system parameter and improving the performance characteristics of the channel [12].
 Shaw [13] has introduced the concept that a deterministically chaotic system can generate entropy. The consequences 
of this observation is important for the design of communication equipment when the channel is a chaotic system. For 
one, it implies that information at the receiver about the state of the channel is lost at a mean rate given by the 
Kolmogorov entropy. For another it implies that a channel estimator should be adapted to the mathematical nature of
 the set of non linear equations describing the channel as shown in the previous paragraph. Our studies up to this 
date have shown that this approach is valid in an indoor situation but not in an outdoor one. This might be due to 
the confined geometry one encounters inside buildings and the boundary conditions for the electromagnetic fields 
leading to a low dimensional system of non linear equations giving birth to the observed chaotic behaviour. Our 
studies in this direction are in progress.\\

{\bf Acknowledgements}

We thank James Theiler, Jean-Pierre Eckmann and Reggie Brown for sending us their computer programs and correspondance, 
as well as Halbert White for some unpublished material. C.T. thanks Sunit Lohtia and Bin Tan for their friendly help 
with the manuscript.\\

\vfill
\centerline{\Large\bf Figure Captions}

\begin{itemize}

\item[Fig.\ 1:] Correlation dimensions vs embedding dimension: Full squares are for Rayleigh distributed points;
 full triangles are for handlimited Rayleigh distributed points. Full diamonds are for experimental results in 
the XPA case whereas open diamonds correspond to the CPA case and open squares to the APA case.

\item[Fig.\ 2:]Liapunov exponent spectrum from Eckmann et al. [7] method versus embedding dimension for the APA
 data (since APA data consist of the largest number of points, 6000).

\item[Fig.\ 3:]Liapunov exponent spectrum from Brown et al. method [8] for the same data as those of Fig.2. 
A linear tangent mapping is used to fit the dynamics. A very similar spectrum is obtained for a second order polynomial.

\item[Fig.\ 4:]Spectrum of generalised dimensions $f(\alpha)$ versus a for the APA data used in fig.2. The
 value at the maximum of $f(\alpha)$ corresponding to the Hausdorff dimension of the strange attractor
 agrees with the minimum bound obtained from fig.1 and with the Kaplan-Yorke conjecture (see text). 
 The spectrum is obtained through embedding in 10 dimensions. This happens to be enough, given the values obtained for
 the various generalised dimensions.

\item[Fig.\ 5:]Measured envelope (APA data used in Fig.2 continuous curve) and its one step prediction (dashed curve)
 from the Neural Network fit to the five dimensional map F (eq.2). The training is over the first 200 first points.

\item[Fig.\ 6:]Normalised prediction error versus embedding. Starting from the same initial weights, we trained 
the neural network, for a given embedding, over the first 200 points with a Marquardt minimisation standard
 deviation parameter equal to 0.01 and total number of 150 iterations. Once, the parameters at the minimum 
error are found we made a one-step ahead prediction over the next 300 points and calculated the resulting 
squared error divided by the total of points. One sees a minimum for an embedding dimension around 5 or 6.
 For an embedding equal to 7 or larger, a large error or no convergence (null pivot encountered in the least 
square error matrices) were observed.

\end{itemize}

\end{document}